\documentclass[preprint2]{aastex}          %% The manuscript based on AASTeX v5.x
\usepackage{graphicx}
\usepackage{units}
\usepackage{color}
\usepackage[bookmarks=true,bookmarksnumbered=false,bookmarksopen=false,
 breaklinks=true,pdfborder={0 0 0},pdfborderstyle={},backref=false,colorlinks=true]
 {hyperref}
\hypersetup{pdftitle={Electromagnetic ion-cyclotron instability in a dusty plasma with 
product-bi-kappa distributions for the plasma particles},
 pdfauthor={M. S. dos Santos, L. F. Ziebell, R. Gaelzer},
 linkcolor=blue, citecolor=blue, urlcolor=blue, pdfnewwindow=true}

\usepackage{amsmath,amsfonts,amssymb}

\begin{document}
%% Article title
%
\title{Electromagnetic ion-cyclotron instability in a dusty plasma with 
product-bi-kappa distributions for the plasma particles
}

%% Running heads
\shorttitle{IC instability in a dusty plasma}
\shortauthors{dos Santos et al.}

%% Author and Affilations
\author{M. S. dos Santos\altaffilmark{1,2}} 
\author{L. F. Ziebell\altaffilmark{1}} 
\and 
\author{R. Gaelzer\altaffilmark{1}} 
%\and 
%\author{\altaffilmark{}}
%\affil{}
%\email{} %% non-output

%% Alternate Affilations
\altaffiltext{1}{Instituto de F\'{\i}sica, Universidade Federal do Rio Grande
do Sul, 91501-970, Porto Alegre, RS, Brasil}
\altaffiltext{2}{Instituto Federal de Educa\c{c}\~ao, Ci\^encia e Tecnologia
Farroupilha, 98590-000, Santo Augusto, RS, Brasil}
%\altaffiltext{2}{}
%\altaffiltext{3}{}

%% Abstract
%\begin{abstract}
\begin{abstract}
We study the dispersion relation for parallel propagating ion-cyclotron (IC)
waves in a dusty plasma, 
considering situations where the velocity
dispersion along perpendicular direction is greater than along the parallel
direction, and considering the use of product-bi-kappa (PBK) velocity 
distributions for the plasma particles.
The results obtained by numerical solution of the dispersion relation, 
in a case with isotropic Maxwellian distributions for 
electrons and PBK distribution for ions, show the
occurrence of the electromagnetic ion-cyclotron instability (EMIC), and show
that the decrease in the kappa indexes of the PBK ion distribution leads to 
significant increase in the magnitude of the growth rates
and in the range of wavenumber for which the instability occurs. 
On the other hand, for anisotropic Maxwellian distribution for ions and 
PBK distribution for electrons, the decrease of the kappa index in the PBK
electron distribution 
contributes to reduce the growth rate of the EMIC instability, 
but the reduction effect is less pronounced than the increase
obtained with ion PBK distribution with the same $\kappa$ index.
The results obtained also show that, as
a general rule, the presence of a dust population contributes to reduce
the instability in magnitude of the growth rates 
and range, but that in the case of PBK ion
distribution with small kappa indexes the instability may continue
to occur for dust populations which would eliminate completely the instability
in the case of bi-Maxwellian ion distributions. It has also been seen that the
anisotropy due to the kappa indexes in the ion PBK distribution is not
so efficient in producing the EMIC instability as the ratio of perpendicular
and parallel ion temperatures, for equivalent value of the effective 
temperature.
\end{abstract}

\section{Introduction}
\label{sec:introduction}

Observations have consistently shown that ions as well as electrons
in the solar wind may have non thermal velocity distributions
which feature characteristic power-law tails
\citep{PilippMMMRS87a,PilippMMMRS87b,MaksimovicPL97,%
MaksimovicZCISLMMSLE05,MarschAT04,Marsch06,HapgoodPDD11}. 
These distributions with
power-law tails can be mathematically described by functions which are 
generically known as kappa distributions, whose introduction for the
description of velocity distributions in space environments is usually
attributed to \cite{Vasyliunas68}. Two different basic
forms of kappa distributions are used in the literature, which are
the kappa distribution as defined in \cite{SummersT91} and the kappa
distribution as defined in \cite{Leubner02}. Anisotropic forms of
kappa distributions can also be defined, either based upon the functions
defined in \cite{SummersT91} or based upon the functions defined in
\cite{Leubner02}, and may be useful for the discussion
of instabilities in space environments. These anisotropic forms of kappa
distributions are usually known as bi-kappa (BK) distributions, when
featuring isotropic kappa index and temperature anisotropy, or as 
product-bi-kappa (PBK) distributions,
when the anisotropy may also occur in the kappa indexes. Another form
of anisotropic kappa distribution is the kappa-Maxwellian distribution, 
characterized by a kappa form along parallel direction and a
Maxwellian distribution along perpendicular directions
\citep{HellbergM02,CattaertHM07}.   
Kappa distributions, either for ions or
for electrons, isotropic or anisotropic,
may lead to significant modifications in the dispersion 
relations of electromagnetic waves and on the growth rates of instabilities
in plasmas \citep{PierrardL10,LazarP09a,LazarP09b,LazarPS11b,Lazar12,LazarPPS12,
LazarP14,LazarPF15,pl:SantosZG14,pl:SantosZG15}.

Bi-kappa distributions have 
been used, for instance, as part of tools to model observed
features of plasmas in the Jupiter environment \citep{MoncuquetBM-V02,%
AndreF08,ImaiLMBHIT15}. Product-bi-kappa distributions are more flexible
than bi-Maxwellians and BK distributions, since their non thermal features can
be modeled by anisotropic temperatures and by anisotropic kappa indexes, and
have already been used as a modeling tool in \cite{LazarPPS12}, and
also in a discussion about the instability 
limits for the Weibel instability, in \cite{LazarSP10}.
We are not aware of other works using PBK distributions to actually fit 
results 
of observations, but it is conceivable that this type of distribution 
can replace with advantage
the less flexible BK distributions or the combinations of bi-Maxwellians which
have been used in past data analysis. It is therefore useful to investigate
the effect of PBK distributions on waves and instabilities, in order to
understand the differences between wave properties in the presence of PBK 
distributions and in the presence of BK or bi-Maxwellian distributions.

In a recent paper we have discussed the dispersion relation and the growth or
damping rates of low frequency electromagnetic waves 
propagating along an ambient magnetic field, 
in a plasma in which electrons and ions can be described by 
PBK distributions, assuming that the plasma may also contain a small 
population of dust particles \citep{pl:SantosZG16}. 
In such a work, we have investigated wave pertaining to the whistler branch
of the dispersion curves.
The growth of the low-frequency waves in this branch 
constitute the so-called ion-firehose instability, a type of
instability which may occur when there is anisotropy 
in the velocity distribution of the ions, caused either
by the occurrence of perpendicular ion temperature smaller than parallel 
temperatures ($T_{i\perp}<T_{i\parallel}$), 
or by anisotropy between the perpendicular and parallel
kappa indexes which characterize a ion PBK distribution.
The work appearing in \cite{pl:SantosZG16}
was closely related to the work appearing in two other relatively recent papers
dedicated to low frequency electromagnetic waves, one that
discussed the ion-firehose instability in a plasma without dust
\citep{pl:SantosZG14}, and another which discussed the
EMIC instability (which occurs when 
$T_{i\perp}>T_{i\parallel}$) \citep{pl:SantosZG15}, also in a plasma without
dust.

The possibility of presence of dust in the plasma, taken into account in
Ref. \cite{pl:SantosZG16}, was motivated by observations which 
have shown that the plasma in the solar wind may contain a dust
population \citep{MannKBTGMLMMGL04,Marsch06,Schwenn06,%
Mann2008,KrugerLAG07,GrunZFG85,IshimotoM98,Meyer-Vernetetal09,Mann2010,
KruegerSGS15}. In fact,
observations of the inner region of the solar system, made 
with satellites like Ulysses, Galileo, or Cassini,
have shown the existence of a population of small dust particles, with sizes 
which range from the nanometric to the micrometric size. The number density
of dust particles has been seen to be much smaller than that of the solar wind 
particles, but it is estimated that the
total mass of dust particles in the inner region of the solar system 
may be about the same order of magnitude as the mass of plasma particles. 
Dust has also been observed as 
a significant element in some specific environments, like in the 
neighborhood of comets \citep{MendisH13}, in planetary rings \citep{Goertz89},
and in the vicinity of Jupiter's satellites \citep{Grunetal97}. 
Dust may therefore play relevant role in the
dynamics of plasma in the solar wind. 
The presence of dust affects the dispersion properties of plasmas,
leading to modifications in the properties of plasma waves in 
comparison with those in a dustless plasma, and also to the occurrence of
new modes, associated to the dynamics of the dust \citep{Shukla1992,Rao1993b}.

In the present paper we return to the study of effects due to form of the
velocity distribution functions, particularly the effect of PBK distributions, 
considering now waves in the IC branch of the 
dispersion relation of low frequency electromagnetic waves,
in a dusty plasma. We concentrate on the study of 
parallel propagating waves,
since this is the condition that leads to the 
maximum growth rate for the EMIC instability
\citep{DavidsonO75,GaryMFF76,Gary05}.
The investigation is complementary to that appearing
in \cite{pl:SantosZG16}, which considered waves in the whistler
branch, and to that appearing in \cite{pl:SantosZG15}, which studied
IC waves in a dustless plasma. 

The paper is organized as follows:
In section \ref{sec:theoretical-formulation} we briefly describe the
theoretical formulation which leads to the dispersion relation for 
electromagnetic waves propagating parallel to the ambient magnetic field in 
a dusty plasma, considering different forms of the velocity distribution for 
plasma particles, either PBK or bi-Maxwellian. The discussion which is 
presented is very brief and dedicated to the presentation and definition of
quantities and expressions used in the present paper, with references to
the literature in what concerns details of the derivation.
Section \ref{sec:numerical-results} is dedicated to presentation and discussion
of results obtained by numerical solution of the dispersion relation, 
considering different combinations of ion and electron distribution functions
and different sets of parameters. Section \ref{sec:conclusions} presents
some final remarks.

\section{Theoretical Formulation}
\label{sec:theoretical-formulation}

For electromagnetic waves which propagate along the ambient magnetic field,
in plasmas with particles described by distribution functions which have
even dependence on the velocity variable which is parallel to the
direction of the magnetic field, 
the dispersion relation is obtained as the solution of 
the following determinant,
\begin{eqnarray}
\label{disprel,1}
\mathrm{det}
\begin{pmatrix}
\varepsilon_{xx}-N_\parallel^2 & \varepsilon_{xy} & 0 \\
-\varepsilon_{xy} & \varepsilon_{xx}-N_\parallel^2 & 0 \\
0 & 0 & \varepsilon_{zz}
\end{pmatrix}=0,
\end{eqnarray}
where it has been taken into account that, for 
$k_\perp\to 0$, $\varepsilon_{yy}=\varepsilon_{xx}$ and 
$\varepsilon_{xz}= \varepsilon_{yz}=0$, and where 
$N_\parallel$ is the parallel component of the
refraction vector, ${\bf N}= c{\bf k}/\omega$.

The determinant given by equation (\ref{disprel,1}) can be 
separated into two minor
determinants, and the dispersion relation of our interest
becomes the following, 
\begin{equation}
\label{disprel,2}
N_\parallel^2=\varepsilon_{xx}\pm i\varepsilon_{xy},
\end{equation}

The components of the dielectric tensor which appear in 
dispersion relation \eqref{disprel,2} in a dusty plasma
can be written as follows
\begin{eqnarray}
\label{epsij}
&&\varepsilon_{xx}=1
+\frac{1}{z^2}\sum_\beta \frac{\omega^2_{p\beta}}{\Omega_i^2}
\frac{1}{n_{\beta 0}}\sum_{m=1}^\infty \left(\frac{q_\perp}{r_\beta}
\right)^{2\left(m-1\right)}\\
&&\quad\times \sum_{n=-m}^m n^2A\left(|n|,m-|n|\right)
J\left(n,m,0;f_{\beta 0}\right)\nonumber
\\
&& \varepsilon_{xy}=-\varepsilon_{yx}
=i\frac{1}{z^2}\sum_\beta \frac{\omega^2_{p\beta}}{\Omega_i^2}
\frac{1}{n_{\beta 0}}\sum_{m=1}^\infty \left(\frac{q_\perp}{r_\beta}
\right)^{2\left(m-1\right)} \nonumber\\
&&\quad\times\sum_{n=-m}^m mnA\left(|n|,m-|n|\right)
J\left(n,m,0;f_{\beta 0}\right)
\nonumber \\
&& \varepsilon_{yy}=1
+\frac{1}{z^2}\sum_\beta \frac{\omega^2_{p\beta}}{\Omega_i^2}
\frac{1}{n_{\beta 0}}\sum_{m=1}^\infty \left(\frac{q_\perp}{r_\beta}
\right)^{2\left(m-1\right)} \nonumber\\
&&\quad\times \sum_{n=-m}^m B\left(|n|,m-|n|\right)
J\left(n,m,0;f_{\beta 0}\right),
\nonumber
\end{eqnarray}
where
\begin{eqnarray*}
\label{integral J}
&&J(n,m,h;f_{\beta 0}) \equiv z \int d^3 u
 \frac{u_\parallel^h u_\perp^{2\left(m-1\right)}u_\perp
}
{z-nr_\beta - q_\parallel u_\parallel +i \tilde{\nu}^{0}_{\beta d}(u)}\\
&&\quad\times \left[\left(1-\frac{q_\parallel}{z} 
u_\parallel \right) \frac{\partial f_{\beta 0}}
{\partial u_\perp}+\frac{q_\parallel}{z} u_\perp \frac{\partial f_{\beta 0}}
{\partial u_\parallel} \right]
\end{eqnarray*}
and where $A(n,m)$ and $B(n,m)$ are numerical coefficients. 
In equations
\ref{epsij}, non dimensional variables have been used, defined
$z= \omega/\Omega_i$, $r_\beta=\Omega_\beta/\Omega_i$,
$\tilde{\nu}^{0}_{\beta d}(u)=\nu^{0}_{\beta d}(u)/\Omega_i$, 
$q_\perp= k_\perp v_A/\Omega_i$,
$q_\parallel= k_\parallel v_A/\Omega_i$, ${\bf u}={\bf v}/v_A$. The quantity
$\omega_{p\beta}=\sqrt{4\pi n_{\beta 0}q_\beta^2/m_\beta}$ is the plasma
frequency for species $\beta$, $\Omega_\beta=q_\beta B_0/(m_\beta c)$
is the cyclotron angular frequency for species $\beta$,
$v_A=B_0/\sqrt{4 \pi n_{i0}m_i}$ is the Alfv\'en velocity, and
$\nu^{0}_{\beta d}(u)$ is the frequency of inelastic collisions between dust
particles and plasma particles of species $\beta$. 
As it is known, the 
presence of an imaginary term in the denominator of the
velocity integrals which appear in the components of the dielectric
tensor, due to collisional charging of dust particles, leads to collisional 
damping of waves, in addition to the collisionless 
damping \cite{TsytovichAB2002,pl:deJuliSZJ05,pl:SchneiderZJJ06}.

In terms of non dimensional variables, the frequency of inelastic dust-plasma
collisions may be written as follows, 
\begin{eqnarray}
\label{collfreq}
&&\nu_{\beta 0}^{j0}(u)=\frac{\pi a^2_j n^j_{d0} v_A}{u}
\left(u^2- \frac{2q_{d0}q_\beta}{m_\beta v_A^2a}\right)\\
&&\quad\times
{\rm H}\left(u^2- \frac{2q_{d0}q_\beta}{m_\beta v_A^2a}\right), \nonumber
\end{eqnarray}
where $q_{d0}$ is the equilibrium value of the dust charge. 
It is assumed
that the dust charge is acquired via inelastic collisions between plasma
particles and the dust grains, and it is therefore convenient
to write $q_{d0}=-eZ_{d0}$, where $Z_{d0}$ is the dust charge number and 
$|e|$ is the absolute value of the charge of and electron. 
The minus sign
occurs because the charge acquired via collisional
processes is negative.
In the derivation of equations \eqref{epsij}, 
it was assumed for simplicity that 
all dust particles are spherical, with radius $a$.
Details of the derivation of equations \eqref{epsij} can be obtained in
\cite{pl:ZiebellSJG08}. 

Using the expressions for the components $\varepsilon_{ij}$ given by
equations \eqref{epsij}, the dispersion
relation can be written in a very simple form, depending on the integral
quantities $J(s,m,h;f_{\beta 0})$,
\begin{equation}
\label{disprel}
N_\parallel^2=1+\frac{1}{2z^2}\sum_\beta \frac{\omega^2_{p\beta}}{\Omega_i^2}
\frac{1}{n_{\beta 0}}J\left(s,1,0;f_{\beta 0}\right)
\end{equation}
with $s=\pm1$. The dependence of the dispersion relation 
on the velocity distribution functions of the
plasma particles appears only in the integral quantities
$J(s,m,h;f_{\beta 0})$. For evaluation of these integral quantities, we
utilize an approximation, which is to use the average value of the
inelastic collision frequency, obtained as follows,
\begin{equation}
\label{nubeta,average}
\nu_\beta \equiv \frac{1}{n_{\beta 0}}\int d^3u \nu_{\beta d}^0(u)
f_{\beta 0}(u),
\end{equation}
instead of the velocity dependent form $\nu_{\beta d}^0(u)$.

For description of the plasma particles, we consider that their velocity 
distributions can be either
PBK distributions or bi-Maxwellian distributions. The PBK distribution
to be used is based on the kappa distribution as
defined in \cite{Leubner02}, and can be written as follows,
\begin{eqnarray}\label{PBK}
&&f_{\beta,\kappa}(u_\parallel,u_\perp)=\frac{n_{\beta0}}{\pi^{3/2}
\kappa_{\beta\perp}\kappa_{\beta\parallel}^{1/2}u_{\beta\perp}^2
u_{\beta\parallel}}\\
&&\quad\times 
\frac{\Gamma(\kappa_{\beta\perp})\Gamma(\kappa_{\beta\parallel})}
{\Gamma(\kappa_{\beta\perp}-1)\Gamma(\kappa_{\beta\parallel}-1/2)} \nonumber \\
&&\quad\times
\left(1+\frac{u_\perp^2}{\kappa_{\beta\perp} u_{\beta\perp}^2}
\right)^{-\kappa_{\beta\perp}}
\left(1+\frac{u_\parallel^2}{\kappa_{\beta\parallel} u_{\beta\parallel}^2}
\right)^{-\kappa_{\beta\parallel}}, \nonumber
\end{eqnarray}
while the bi-Maxwellian velocity distribution is given in dimensionless
variables by the well known expression,
\begin{eqnarray}
\label{bimax}
f_{\beta M}(u_\parallel, u_\perp)&=&
\frac{n_{\beta 0}}{\pi^{3/2}u_{\beta\perp}^2u_{\beta\parallel}}
e^{-u^2_\perp/u_{\beta\perp}^2} e^{-u^2_\parallel/u_{\beta\parallel}^2}.
\end{eqnarray}

In equation \eqref{PBK}, and in equation \eqref{bimax} 
as well, we have used
the non dimensional thermal velocities for particles of species $\beta$, 
defined as
$$
u_{\beta\perp}^2=\frac{2T_{\beta\perp}}{m_\beta v_A^2},\quad
u_{\beta\parallel}^2=\frac{2T_{\beta\parallel}}{m_\beta v_A^2},
$$
where $T_{\beta\perp}$ and $T_{\beta\parallel}$ denote respectively the 
perpendicular and parallel temperatures for particles of species $\beta$. The
anisotropy of temperatures for a given species is usually denoted by the ratio
$T_{\beta\perp}/T_{\beta\parallel}$.

Equation \eqref{PBK} is the form of PBK distribution which has been
used in a previous study made on the EMIC instability for dustless plasmas 
\citep{pl:SantosZG15}, and also in studies made on waves in the whistler branch,
in dustless and in dusty plasmas \citep{pl:SantosZG14,pl:SantosZG16},
respectively.

As can be easily verified from equation \eqref{PBK}, a 
PBK distribution has two sources of anisotropy between
perpendicular and parallel directions, the difference in the 
temperature variables $T_{\beta\perp}$ and $T_{\beta\parallel}$, and the 
difference in the kappa indexes $\kappa_{\beta\perp}$
and $\kappa_{\beta\parallel}$. Effective temperatures along perpendicular and 
parallel directions can be defined, and in
energy units are given as follows,
\begin{eqnarray}			
\label{paralleltemp}				
\theta_{\beta \perp}=m_\beta\int d^3v\,\frac{v_\perp^2}{2} f_{\beta 0}
=\frac{\kappa_{\beta\perp}}{\kappa_{\beta\perp}-2}T_{\beta\perp},\\
\theta_{\beta \parallel}=m_\beta\int d^3v\,v_\parallel^2 f_{\beta 0}
=\frac{\kappa_\parallel}{(\kappa_\parallel-3/2)}T_{\beta \parallel}.\nonumber
\end{eqnarray}

It is seen that the effective temperatures $\theta_{\beta\perp}$ and
$\theta_{\beta\parallel}$ are larger than the corresponding temperatures
$T_{\beta\perp}$ and $T_{\beta\parallel}$. This feature is characteristic
of distributions based on the kappa distribution as defined in
\citep{Leubner02}. Kappa distributions as defined in
\cite{SummersT91}, which are not considered in the present paper,
feature effective temperatures which are equal to the respective temperatures
$T_{\beta\perp}$ and $T_{\beta\parallel}$
\citep{LivadiotisM13,LivadiotisM13b,Livadiotis2015}. 

Using $f_{\beta 0}$ as the distribution given by equation (\ref{PBK}), 
with use of the
approximation given by equation \eqref{nubeta,average}, the integral $J$
appearing in the dispersion relation, given by equation \eqref{disprel},
becomes as follows \citep{pl:GalvaoZGJ11}
\begin{eqnarray}
\label{J,PBK}
&&
J(s,1,0;f_{\beta 0})
=n_{\beta 0} \frac{2\kappa_{\beta\perp}}{\kappa_{\beta\perp} 
-2} \biggl[-\frac{\kappa_{\beta\perp} -2}{\kappa_{\beta\perp}}\\
&&\quad
+\frac{u^2_{\beta \perp}}
{u^2_{\beta \parallel}}\frac{\kappa_{\beta\parallel} -1/2}
{\kappa_{\beta\parallel}}
+\left(\zeta_\beta^0 - \hat{\zeta}^s_\beta \right)\frac{\kappa_{\beta\perp} -2}
{\kappa_{\beta\perp}}Z^{\left(0\right)}_{\kappa_{\beta\parallel}}
({\zeta}^s_\beta)\nonumber \\
&&\quad 
+\frac{u^2_{\beta \perp}}{u^2_{\beta \parallel}}{\zeta}^s_{\beta}
Z^{\left(1 \right)}_{\kappa_{\beta\parallel}}({\zeta}^s_\beta)
\biggr], \nonumber
\end{eqnarray}
where
\begin{eqnarray}
\label{zeta0,zetan}
&& \zeta_\beta^0=\frac{z}{q_\parallel u_{\beta \parallel}},\quad
\zeta_\beta^s=\frac{z-sr_\beta +i\tilde{\nu}_\beta}
{q_\parallel u_{\beta \parallel}},
\end{eqnarray}
\begin{eqnarray}
\label{Z zero}
&&Z_{\kappa_{\beta\parallel}}^{\left(0 \right)}\left(\xi \right)
=i \frac{\kappa_{\beta\parallel} -1/2}{\kappa_{\beta\parallel}^{3/2}}\\
&&\quad\times  
{}_2F_1\left[1,2\kappa_{\beta\parallel}, \kappa_{\beta\parallel}+1;
\frac{1}{2}\left(1+\frac{i\xi}{\kappa_{\beta\parallel}^{1/2}}\right) \right],
\nonumber\\
&&\quad \kappa_{\beta\parallel}>-1/2 \nonumber
\end{eqnarray}
\begin{eqnarray}
\label{Z um}
&&Z_{\kappa_{\beta\parallel}}^{\left(1 \right)}\left(\xi \right)
=i \frac{\left(\kappa_{\beta\parallel} -1/2\right)
\left(\kappa_{\beta\parallel}+1/2\right)}
{\kappa_{\beta\parallel}^{3/2}\left(\kappa_{\beta\parallel}+1
\right)} \\
&&\quad \times  
{}_2F_1\left[1,2\kappa_{\beta\parallel}+2, \kappa_{\beta\parallel}+2;
\frac{1}{2}\left(1+\frac{i\xi}{\kappa_{\beta\parallel}^{1/2}}\right) \right], 
\nonumber\\
&&\quad\kappa_{\beta\parallel}>-3/2, \nonumber
\end{eqnarray}
and where ${}_2 F_1$ is the hypergeometric Gauss function.

Using $f_{\beta 0}$ as the bi-Maxwellian distribution function
given by equation \eqref{bimax}, 
and using the approximation given by equation
\eqref{nubeta,average}, the $J$ integral which has to be used in the
dispersion relation becomes the following
\begin{eqnarray}
\label{J,bimax}
&&J(s,1,0;f_{\beta 0})= 2\,n_{\beta 0}
\Biggl\{ \zeta_\beta^0 Z({\zeta}_\beta^n)\\
&&\quad -\left(1-\frac{u_{\beta\perp}^2}{u_{\beta\parallel}^2}\right)
\left[ 1 + {\zeta}_\beta^n Z({\zeta}_\beta^n) \right]
\Biggr\},\nonumber
\end{eqnarray}
where $Z$ is the well-known plasma dispersion function \citep{FriedConte61},
\begin{equation}
\label{funcao de dispersao do plasma}
Z(\zeta)=\frac{1}{\sqrt{\pi}} \int_{-\infty}^{+\infty} dt
\frac{e^{-t^2}}{t-\zeta}
\end{equation}

A timely comment is that equation (\ref{J,PBK}), for 
$\kappa_{\beta\parallel} \rightarrow \infty$ and $\kappa_{\beta\perp} 
\rightarrow \infty$, becomes the same as equation (\ref{J,bimax}).

\section{Numerical  Analysis}
\label{sec:numerical-results}

For the numerical analysis about the effect of dust on the dispersion relation
of IC waves, and on the growth rate of the EMIC instability,
we consider the same parameters which we have used in our
previous study about the IC waves, which was made without considering
the presence of dust \citep{pl:SantosZG15}, and add the dust population. 
For the non dimensional
parameters $v_A/c$ and $\beta_i$ we assume the values 
$v_A/c= 1.0\times10^{-4}$ and $\beta_i= 2.0$ (except for one figure, where
we take $\beta_i=0.2$), for the ion charge number we
take $Z_i=1.0$, and for the ion mass we assume $m_i=m_p$, the mass of a proton.
Moreover, since we are interested in an instability generated by anisotropy 
in the ion distribution, we consider in all cases to be studied that the 
electron temperature is isotropic and is
the same as the ion parallel temperature, 
$T_{e}=T_{i\parallel}$. With this choice of parameters, the results to
be obtained in the present analysis, in the limit of vanishing dust population,
can be directly compared with the results obtained in  
\cite{pl:SantosZG15}. 
When a dust population is taken into account, we assume micrometric
particles, with radius $a=\unit[1.0\times 10^{-4}]{cm}$. Moreover, 
in the ensuing numerical analysis, the number density of the dust population 
is given in terms of the number density of the ion population, which is another
parameter which has to be assumed. We take the ion number density as
$n_{i0}=\unit[1.0\times 10^9]{cm^{-3}}$, value which can be representative
of outbursts in carbon-rich stars, environment where effects associated to 
the presence of dust are expected to be very significant \citep{TsytovichMT04}. 
In fact, it was even argued that in these environments the 
conditions may be such that dust grains can eventually become coupled, forming
dust molecules \citep{TsytovichMT04}. Such strongly coupled dusty plasmas
are not contemplated in our analysis, which is valid for weakly coupled dust.
With the choice made for the ion number density, 
the results to be obtained in the present analysis of dusty plasmas
can be compared, in the limit where the particle distributions
tend to be Maxwellian or bi-Maxwellian, with results obtained in
\cite{pl:deJuliSZG07b}, where the same set of parameters has been 
utilized.

\begin{figure}
\plotone{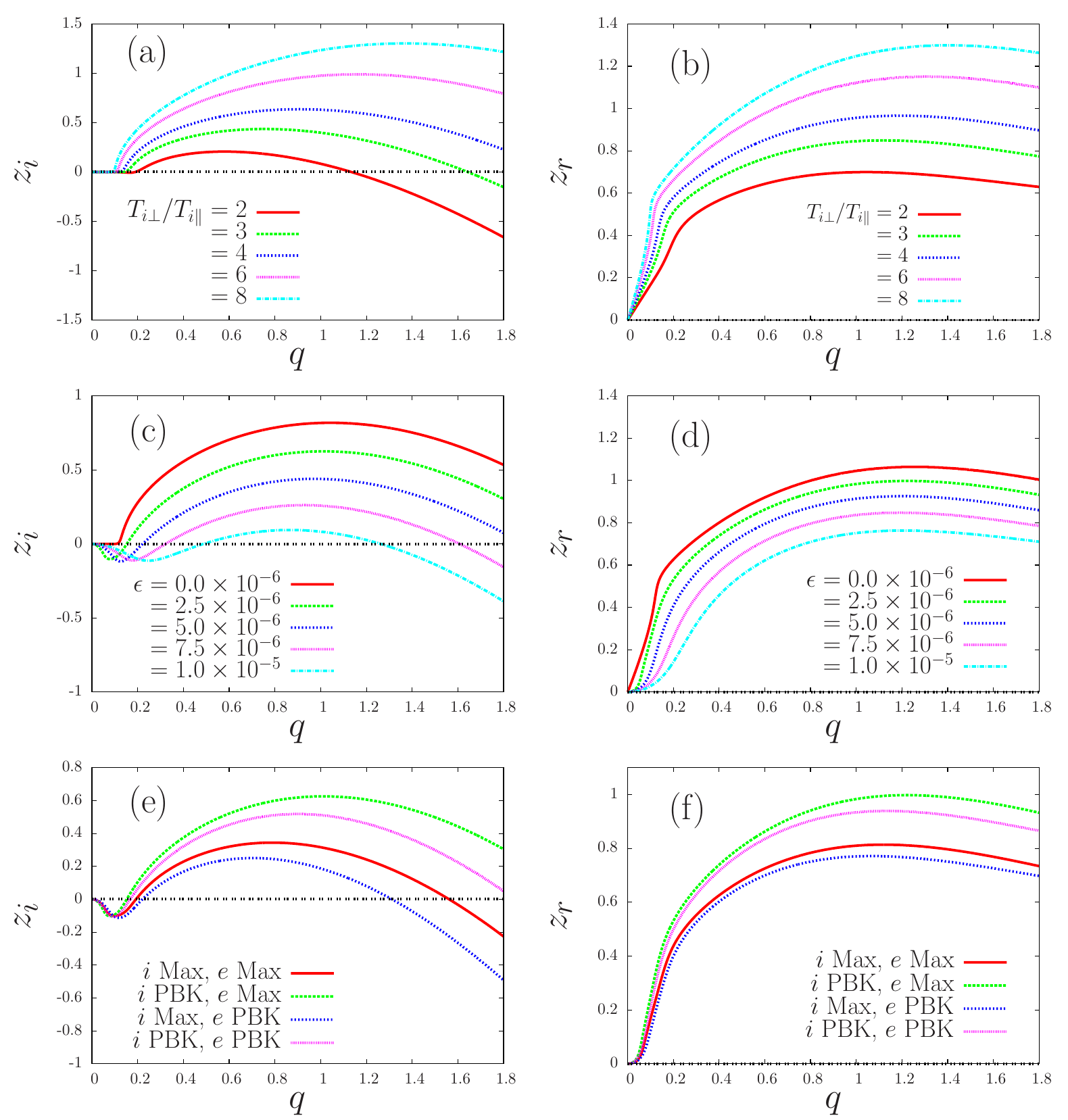}
\caption{Imaginary part ($z_i$, left column) and  
real part ($z_r$, right column),
of the normalized frequency for IC mode waves, vs. normalized 
wave number $q$.
The plasma parameters are
$\beta_i= 2.0$, $v_A/c= 1.0\times 10^{-4}$, 
$n_{i0}=\unit[1.0\times 10^9]{cm^{-3}}$, and
$a=\unit[1.0\times 10^{-4}]{cm}$, with isotropic electron temperature,
$T_e=T_{i\parallel}$.
(a) and (b) PBK distribution for ions, with 
$\kappa_{i\perp}=\kappa_{i\parallel}= 5.0$, and isotropic Maxwellian
for electrons, and several values of the ion temperature ratio 
($T_{i\perp}/T_{i\parallel}= 2$, 3, 4, 6, and 8),
in the absence of dust ($\epsilon=0$);
(c) and (d) same combination of distribution functions as in panels (a) and (b),
with $T_{i\perp}/T_{i\parallel}= 5$, 
and different values of the dust density ratio
($\epsilon=0$, $2.5\times 10^{-6}$, $5.0\times 10^{-6}$, $7.5\times 10^{-6}$,
and $1.0\times 10^{-5}$);
(e) and (f) different combinations of ion and electron distribution functions, 
with $\epsilon=2.5\times 10^{-6}$ and $T_{i\perp}/T_{i\parallel}$= 5.0.
PBK distributions are evaluated 
with $\kappa_{\beta\perp}=\kappa_{\beta\parallel}=5$. 
}
\label{fig1}
\end{figure}

Using these parameters, we solve the dispersion relation considering
different forms of the distribution functions for ions and electrons,
and different values of the number density population of dust particles.
In the figures which follow, the left columns show the
values of the imaginary part of the normalized frequency, $z_i$, and the 
right columns show values of the real part, $z_r$.
Figures \ref{fig1}(a) and \ref{fig1}(b) show the results obtained
considering for the ions a PBK distribution with
$\kappa_{i\perp}=\kappa_{i\parallel}=5.0$, and for the electrons
an isotropic Maxwellian distribution, in the absence of dust ($\epsilon=0.0$).
With the chosen distributions, for the ions the integral
$J$ in the dispersion relation \eqref{disprel} is given by equation
\eqref{J,PBK}, and for the electrons by equation \eqref{J,bimax}.
The individual curves show the results obtained considering different values of 
the ion temperature anisotropy, $T_{i\perp}/T_{i\parallel}$= 2, 3, 4, 6, and 8.
It is seen that the growth rates of the instability, that is, the magnitude
of the values of $z_i>0$, increase with the increase of the temperature ratio.
It is also seen that
the range of values of $q$ where instability occurs increases as well. 
In figures \ref{fig1}(c) and \ref{fig1}(d) we show results obtained considering
the same combination of particle distribution functions used in the case of
figures \ref{fig1}(a) and (b), but assuming $T_{i\perp}/T_{i\parallel}=5$ and
several values of the relative dust population, $\epsilon= n_{d0}/n_{i0}$.
Figure \ref{fig1}(c) shows that the overall effect of the presence
of dust is a reduction of the instability, both in the magnitude of the
growth rates and in the range of unstable wave numbers. It is seen that 
for the combination of distributions which has been considered
the instability will be suppressed if the relative
number density of the dust population becomes somewhat above 
$\epsilon=1.0\times 10^{-5}$.
At the right-hand side, figure \ref{fig1}(d)
shows that the values of $z_r$ decreases with the increase of dust density, 
for all values of wavenumber. 

The influence of different forms of the velocity distribution functions in
a dusty plasma is
discussed in panels (e) and (f) of figure \ref{fig1}, which were obtained
considering $T_{i\perp}/T_{i\parallel}=5$ and $\epsilon= 2.5\times 10^{-6}$.
The red curves in figures \ref{fig1}(e) and \ref{fig1}(f) are obtained for the 
case of bi-Maxwellian ion distribution and isotropic Maxwellian distribution
for the electrons,
The green curves are obtained for the case of PBK ion distribution 
with $\kappa_{i\perp}=\kappa_{i\parallel}=5.0$,
and isotropic Maxwellian distribution for the electrons, and
correspond to the green curves in panels (c) and (d).
The comparison between the green and the red curves in 
panel (f) show that the PBK ion distribution leads to
significant increase in the instability growth rates and on the range of
wave numbers with instability, in comparison with the bi-Maxwellian case.
There is also the appearing of a region of damping for small values of
$q$, which is not present in dustless plasmas.
On the other hand, for bi-Maxwellian ion distribution, and a PBK distribution
for the electrons, with $\kappa_{e\perp}=\kappa_{e\parallel}= 5.0$, the blue 
curves in figures \ref{fig1}(e) and (f) show that the magnitude of the growth
rates is reduced as compared to the case of Maxwellian electron distribution.
If PBK distributions are assumed for both ions and electrons,
with $\kappa_{e\perp}=\kappa_{e\parallel}=5.0$ and
$\kappa_{i\perp}=\kappa_{i\parallel}=5.0$, the magenta curves in panels
(c) and (d) show that the effect is a significant increase of the growth rates
in comparison with the result depicted by the blue curves. However, the
growth rates are smaller than those depicted by the green curves. 
The conclusions are that, for the same kappa indexes, the PBK ion distribution
contributes to increase the growth rate of the EMIC, and that this
effect is more significant than the tendency of decrease in the growth rates,
due to a PBK electron distribution.

\begin{figure}
\plotone{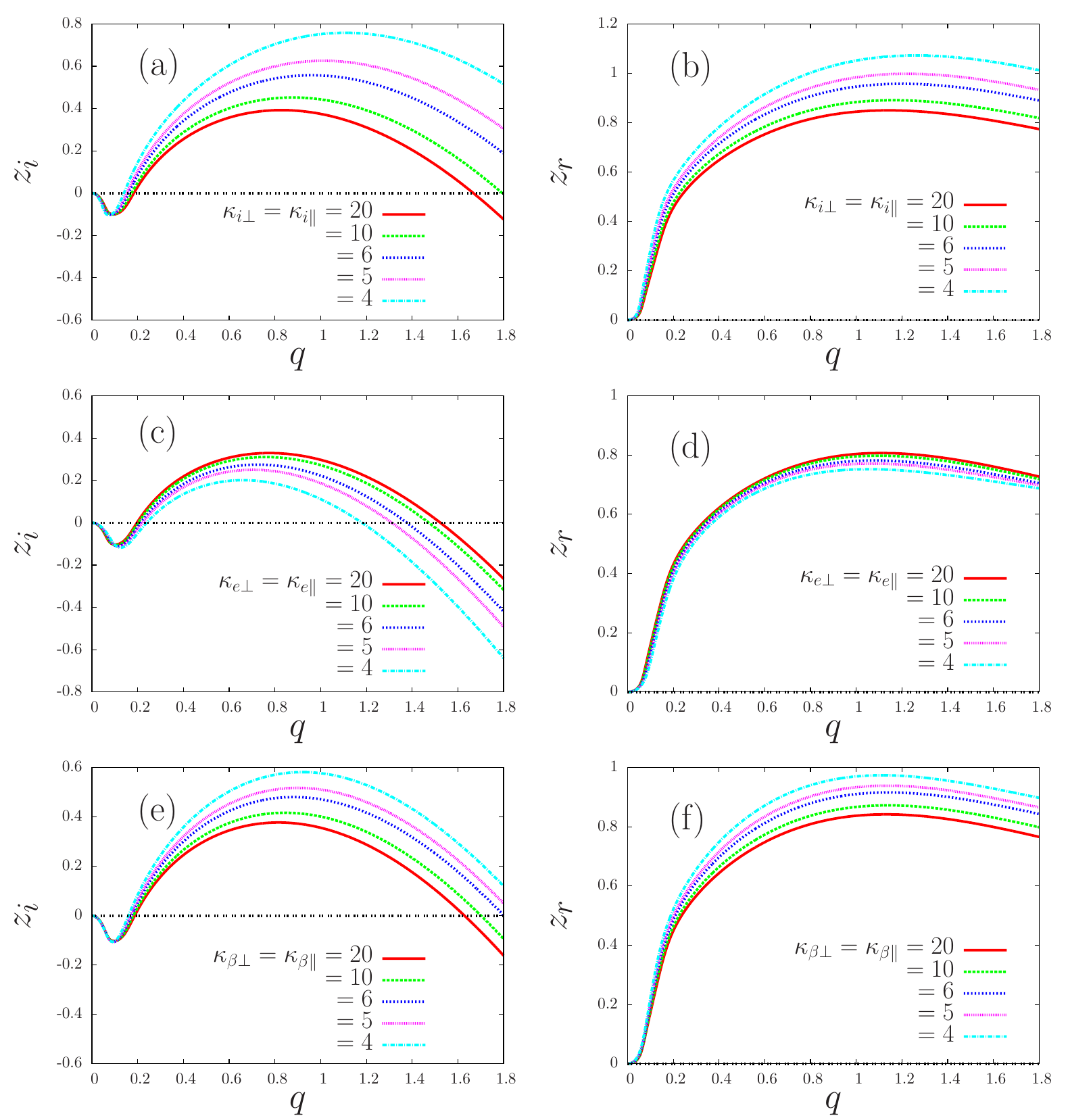}
\caption{Imaginary part ($z_i$, left column) and  
real part ($z_r$, right column),
of the normalized frequency for IC mode waves, vs. normalized 
wave number $q$, for different forms of ion and electron distribution
functions, with several values of the kappa index.
Parameters are $T_{i\perp}/T_{i\parallel}$= 5.0, 
$\epsilon=2.5\times 10^{-6}$, and other parameters as in figure
\protect{\ref{fig1}}.
(a) and (b) PBK distribution with $\kappa_{i\perp}=\kappa_{i\parallel}$ for
ions, with $\kappa$ values 20, 10, 6, 5, and 4, 
and Maxwellian distribution for electrons;
(c) and (d) Bi-Maxwellian distribution for ions and PBK distribution for
electrons, with $\kappa_{e\perp}=\kappa_{e\parallel}$, with 
$\kappa$ values 20, 10, 6, 5, and 4; 
(e) and (f) PBK distribution with $\kappa_{i\perp}=\kappa_{i\parallel}=5$ for
ions and PBK distribution for electrons, with 
$\kappa_{e\perp}=\kappa_{e\parallel}$, and $\kappa$ values 20, 10, 6, 5, and 4.
}
\label{fig2}
\end{figure}

In figure \ref{fig2} we investigate the effect of the value of the kappa
index on the EMIC instability, when ions and/or electrons are
described by PBK distributions, considering
fixed values of ion temperature ratio, $T_{i\perp}/T_{i\parallel}=5.0$, and
dust relative number density, $\epsilon=2.5\times 10^{-6}$, with other
parameters the same as those used in the case of figure \ref{fig1}. 
In figures \ref{fig2}(a) and \ref{fig2}(b) 
we show results obtained in the case that the electrons
are described by a Maxwellian distribution, and the ions are described by a
PBK distribution, considering five values of the kappa indexes,
$\kappa_{i\perp}=\kappa_{i\parallel}=20$, 10, 6, 5, and 4. 
The ratios of effective temperatures are therefore 
$\theta_{i\perp}/\theta_{i\parallel}= 5.14$, 5.31, 5.63, 5.83, and 6.25,
respectively. 
Figure \ref{fig2}(a) shows that for $\kappa_i$ indexes equal 
to 20, the EMIC instability occurs for $0.2\le q\le 1.65$. With decrease of the
kappa indexes to 10, the upper limit of the unstable region is moved to
$q\simeq 1.8$, and the lower limit is decreased from the previous value.
With further decrease of the kappa indexes, the unstable region increases 
considerably, extending well beyond the maximum value of $q$ which is shown
in figure \ref{fig1}. The lower limit of the unstable region continues to
decrease, being $q\simeq 0.13$ for kappa indexes equal to 4. The maximum value
of the growth rate also increases with the decrease of the ion kappa indexes,
being $z_i\simeq 0.38$ at $q\simeq 0.8$, in the case of
$\kappa_{i\perp}=\kappa_{i\parallel}=20$, and 
$z_i\simeq 0.77$ at $q\simeq 1.1$, in the case of
$\kappa_{i\perp}=\kappa_{i\parallel}=4$. 
For the real part of the dispersion relation, figure
\ref{fig2}(b) shows that the effect of reduction of the kappa index in the
ion distribution is and increase of the value of $z_r$, at all wavelengths.

In figure \ref{fig2}(c) and \ref{fig2}(d) we present results obtained from
numerical solution of the dispersion relation in the case of ions described
by a bi-Maxwellian distribution and electrons described by PBK distribution,
with $\kappa_{e\perp}=\kappa_{e\parallel}=20$, 10, 6, 5, and 4. 
Despite the isotropy of electron temperature,
$T_{e\perp}=T_{e\parallel}$, there is an effective electron
anisotropy given by 
$\theta_{e\perp}/\theta_{e\parallel}$, which goes
from 1.03 in the case of kappa indexes equal to 20, to 1.25 in the case of
kappa indexes equal to 4. Figure \ref{fig2}(c) shows that the decrease of the
kappa indexes of the electron distribution is associated to decrease in the 
magnitude of the growth rates 
and range of the EMIC instability, contrarily to what was observed
in the case of decrease of the kappa indexes of the ion distribution, in
figure \ref{fig2}(a). For the real part of the dispersion relation, figure
\ref{fig2}(d) shows that the effect of the kappa index in the electron 
distribution is very small, much less significant than the effect due to the
ion kappa indexes, seen in figure \ref{fig2}(b).

In figure \ref{fig2}(e) and \ref{fig2}(f) we depict results obtained 
considering both ions and electrons described by PBK distributions,
with the same kappa indexes, considering kappa indexes 20, 10, 6, 5, and 4. 
In consonance with the results shown in figures \ref{fig2}(a)
and \ref{fig2}(c), where it is seen that the growth rate of the EMIC 
instability tends to increase with the decrease of the kappa index in the
ion PBK distribution, and decrease with the decrease of the kappa index in 
the electron PBK distribution, the results shown in figure 
\ref{fig2}(e) feature growth rates with values which are in between those 
obtained in panels (a) and (c). For 
instance, in the case of kappa values equal to 5, figure \ref{fig2}(e) shows
that the maximum value of $z_i$ is nearly 0.51, while it is $z_i\simeq 0.6$
in the case of figure \ref{fig2}(a) and $z_i\simeq 0.23$ in figure
\ref{fig2}(c).

\begin{figure}
\plotone{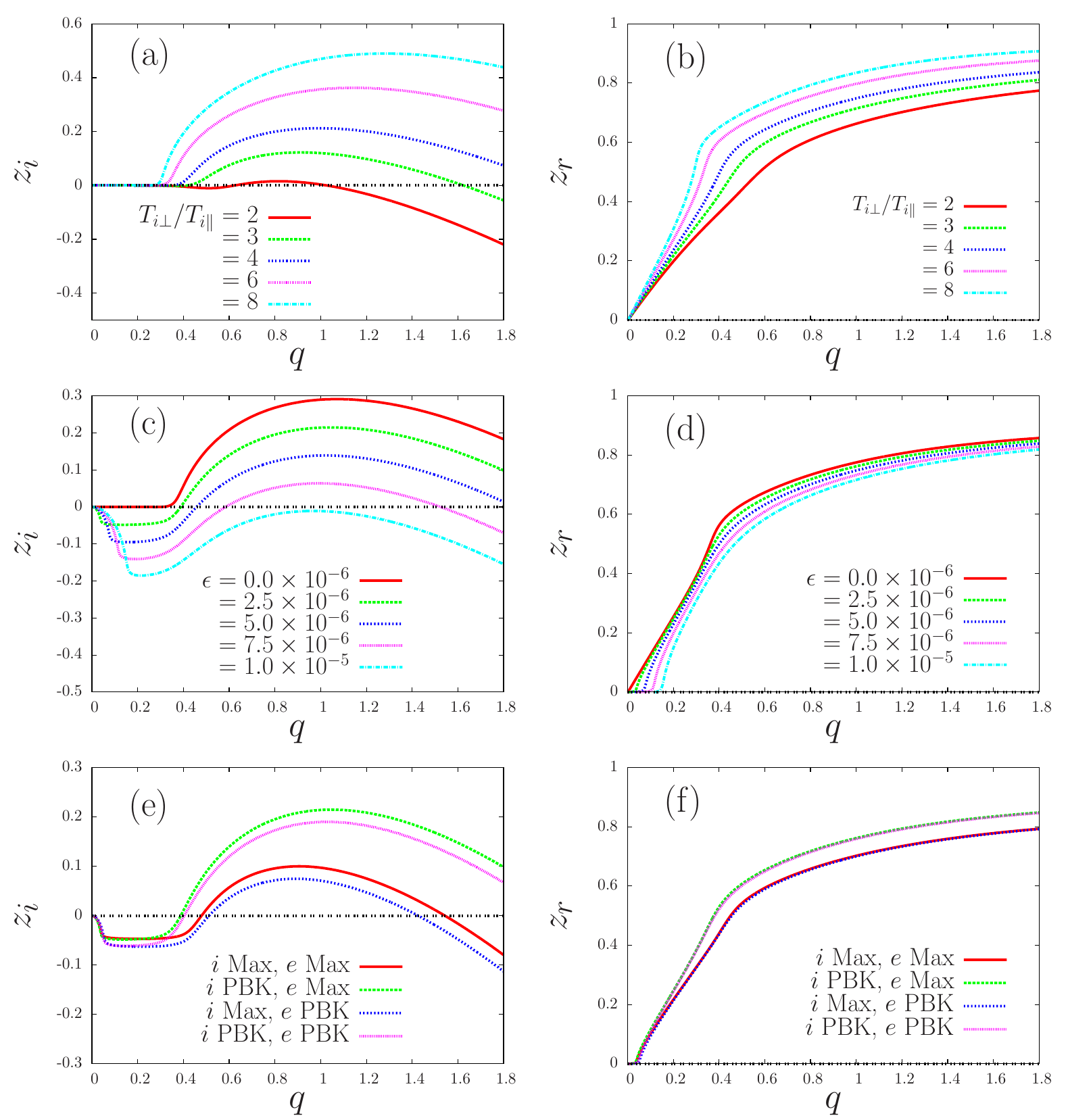}
\caption{Imaginary part ($z_i$, left column) and  
real part ($z_r$, right column),
of the normalized frequency for IC mode waves, vs. normalized 
wave number $q$.
The plasma parameters are $\beta_i= 0.2$, 
and other parameters as in figure \protect{\ref{fig1}}.
(a) and (b) PBK distribution for ions, with 
$\kappa_{i\perp}=\kappa_{i\parallel}= 5.0$, and isotropic Maxwellian
for electrons, and several values of the ion temperature ratio 
($T_{i\perp}/T_{i\parallel}= 2$, 3, 4, 6, and 8),
in the absence of dust ($\epsilon=0$);
(c) and (d) same combination of distribution functions as in panels (a) and (b),
with $T_{i\perp}/T_{i\parallel}= 5$, 
and different values of the dust density ratio
($\epsilon=0$, $2.5\times 10^{-6}$, $5.0\times 10^{-6}$, $7.5\times 10^{-6}$,
and $1.0\times 10^{-5}$);
(e) and (f) different combinations of ion and electron distribution functions, 
with $\epsilon=2.5\times 10^{-6}$ and $T_{i\perp}/T_{i\parallel}$= 5.0.
PBK distributions are evaluated 
with $\kappa_{\beta\perp}=\kappa_{\beta\parallel}=5$. 
}
\label{fig3}
\end{figure}

Figure \ref{fig3} reproduces the analysis made in figure \ref{fig1}, but
considering $\beta_i=0.2$, instead of $\beta=2.0$.
The comments which can be made about the results shown in figure \ref{fig3}(a)
are qualitatively similar to those made about figure \ref{fig1}(a), with the
difference that the onset of the instability in the case of lower $\beta_i$
occurs for larger values of $q$ than in the case of higher $\beta_i$, and that
the instability in the case of lower $\beta_i$ requires slightly larger value
of the ion temperature ratio in order to be started.
Regarding the dependence on the dust density, 
figure \ref{fig3}(c) shows that the 
presence of dust leads to a reduction of the magnitude 
of the growth rates and
in the range of unstable wave numbers, 
as in the case of higher $\beta_i$ presented in figure \ref{fig1}, 
with effectiveness which 
depends upon the velocity distribution functions. In the case of electrons 
with Maxwellian distribution and ions with PBK distribution with
$\kappa_{i\perp}=\kappa_{i\parallel}=5.0$, 
shown in figures \ref{fig3}(c) and \ref{fig3}(d), the 
EMIC instability ceases to exist for $\epsilon\simeq 1.0\times 10^{-5}$.
Panels (e) and (f) show the results of analysis
of the effect of change in the particle distribution functions, for a given
value of dust density ($\epsilon=2.5\times 10^{-6}$). It is seen
in figure \ref{fig3}(e) that the use of a PBK ion distribution leads to 
significant increase in the magnitude of the growth rates and in the range of 
unstable wave numbers, similarly to what has been seen for higher $\beta_i$
in figure \ref{fig1}(e). Figure \ref{fig3}(e) also shows that the use of a 
PBK electron distribution leads to some decrease in the magnitude of the growth
rate, but the effect is less pronounced than obtained in the case of figure
\ref{fig1}(e). It is also seen in figure \ref{fig3}(e) that the damping region
which appears for small $q$ is considerably larger than obtained in the case
of higher $\beta_i$, in figure \ref{fig1}(e).
Regarding the real part, the three panels in the right column of figure 
\ref{fig3} show effects qualitatively similar to those shown in figure
\ref{fig1}, but with less intense in the case of lower $\beta_i$ than in the 
case of higher $\beta_i$.

\begin{figure}
\plotone{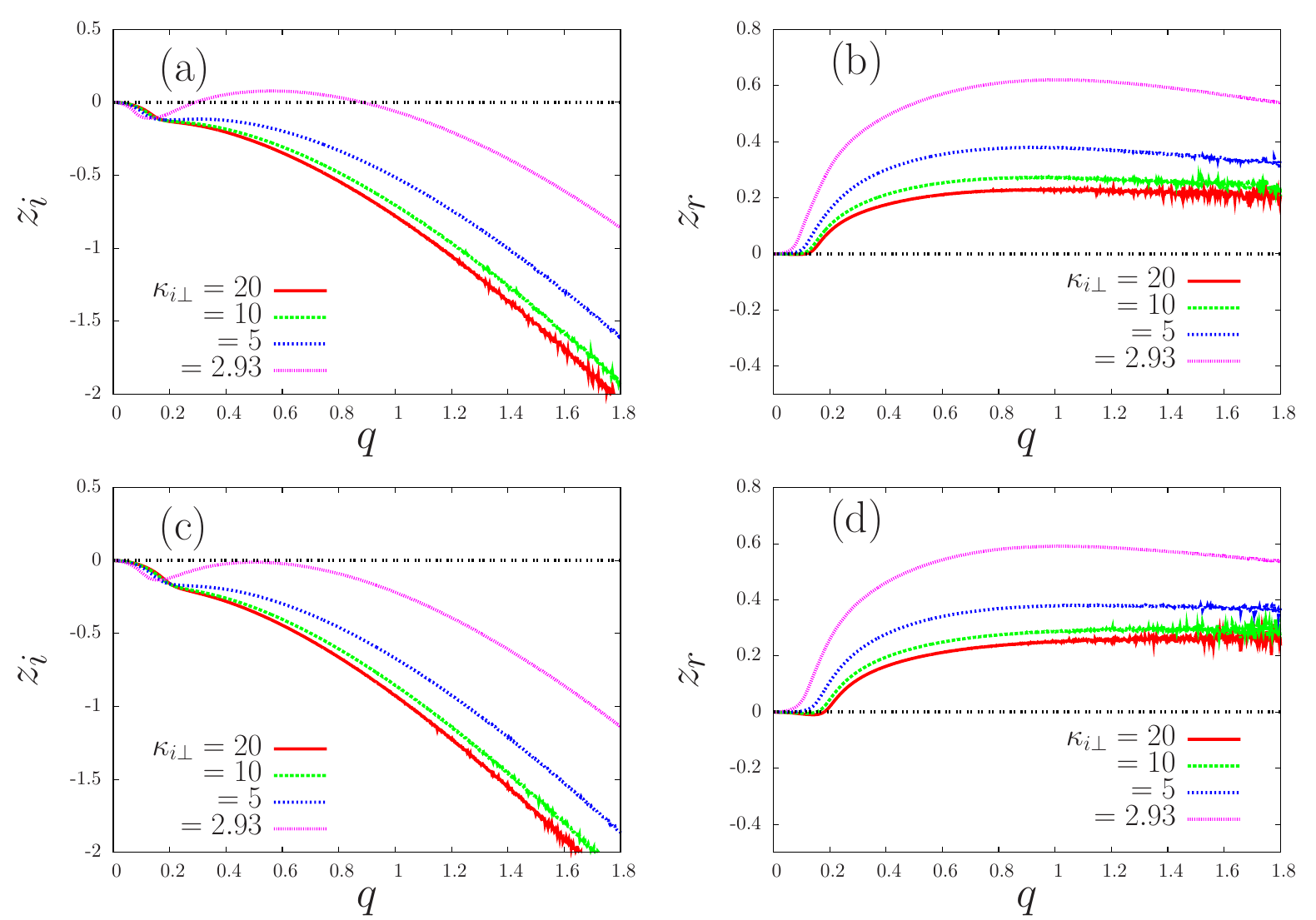}
\caption{Imaginary part ($z_i$, left column) and  
real part ($z_r$, right column),
of the normalized frequency for IC mode waves, vs. normalized 
wave number $q$, for different forms of ion and electron distribution
functions, with isotropic ion temperatures and 
different values of anisotropy in the ion kappa indexes.
Parameters are $\beta_i= 2.0$, $v_A/c= 1.0\times 10^{-4}$,
$n_{i0}=\unit[1.0\times 10^9]{cm^{-3}}$,
$a=\unit[1.0\times 10^{-4}]{cm}$, 
$\epsilon= 2.5\times 10^{-6}$,
and isotropic electron temperature
$T_e= T_{i\parallel}$.
(a) and (b) PBK distribution for ions, with $\kappa_{i\parallel}=30$ and 
several  values of $\kappa_{i\perp}$ (20, 10, 5, and 2.93), 
and isotropic Maxwellian for electrons;
(c) and (d) PBK distribution for ions, with $\kappa_{i\parallel}=30$ and 
several  values of $\kappa_{i\perp}$ (20, 10, 5, and 2.93),
and PBK distribution for electrons, with 
$\kappa_{e\parallel}=\kappa_{e\perp}=5.0$.
}
\label{fig4}
\end{figure}

\begin{figure}
\plotone{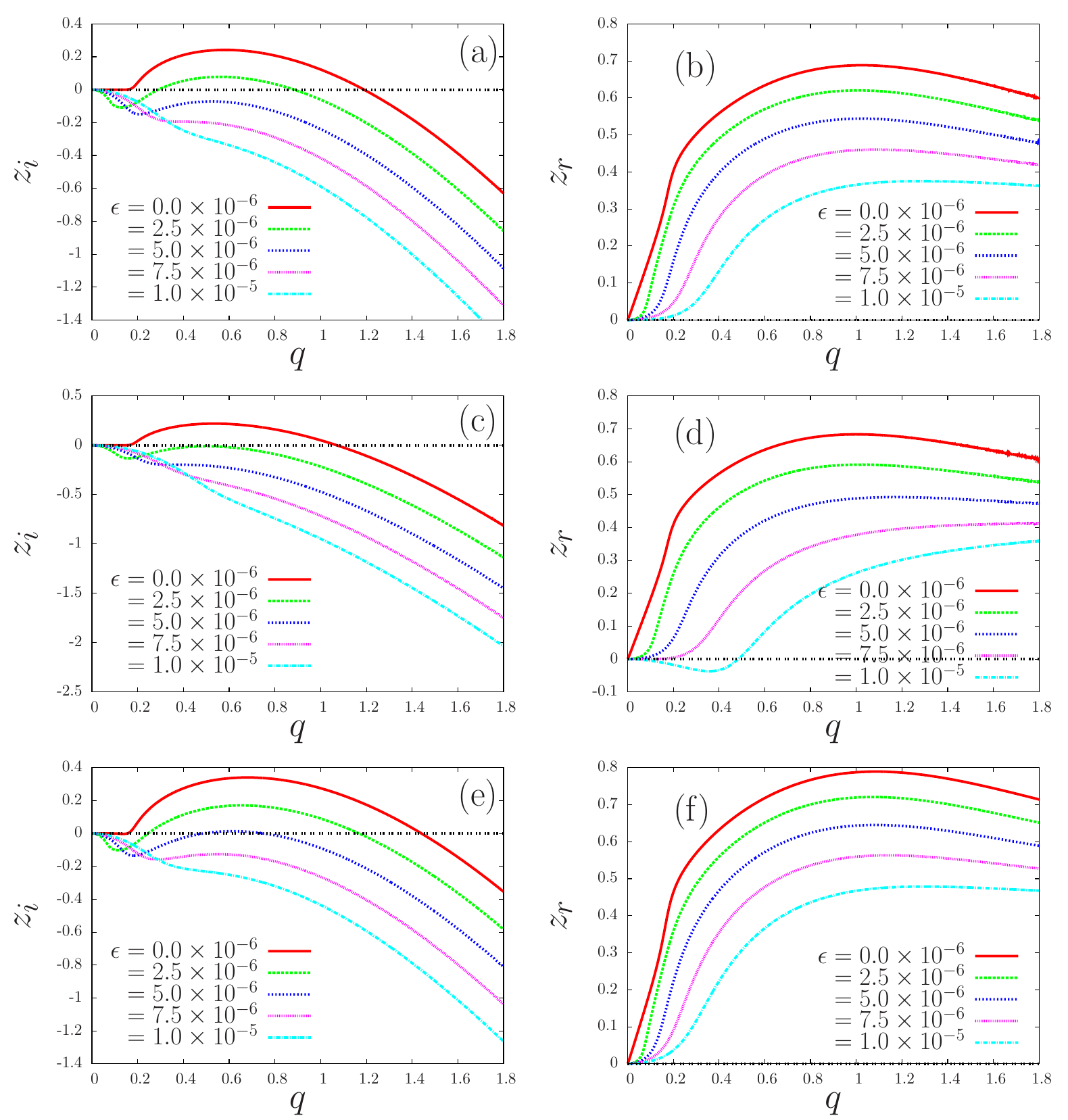}
\caption{Imaginary part ($z_i$, left column) and  
real part ($z_r$, right column),
of the normalized frequency for IC mode waves, vs. normalized 
wave number $q$, for different forms of ion and electron distribution
functions, and several values of the relative dust number density
($\epsilon=0$, $2.5\times 10^{-6}$, $5.0\times 10^{-6}$, $7.5\times 10^{-6}$,
and $1.0\times 10^{-5}$).
Other parameters are $\beta_i= 2.0$, $v_A/c= 1.0\times 10^{-4}$,
$n_{i0}=\unit[1.0\times 10^9]{cm^{-3}}$,
$a=\unit[1.0\times 10^{-4}]{cm}$, and isotropic electron temperature
$T_e= T_{i\parallel}$.
(a) and (b) PBK distribution for ions, with $\kappa_{i\parallel}=30$ and 
$\kappa_{i\perp}=2.93$, with $T_{i\perp}=T_{i\parallel}$,
and isotropic Maxwellian for electrons;
(c) and (d) PBK distribution for ions, with 
$\kappa_{i\parallel}=30$ and $\kappa_{i\perp}=2.93$, 
with $T_{i\perp}=T_{i\parallel}$, and
PBK distribution for electrons, with 
$\kappa_{e\parallel}=\kappa_{e\perp}=5.0$;
(e) and (f) PBK distribution for ions, with $\kappa_{i\parallel}
=\kappa_{i\perp}=5.0$ and $T_{i\perp}/T_{i\parallel}= 2.565$, and
isotropic Maxwellian for electrons. 
}
\label{fig5}
\end{figure}

In figure \ref{fig4} and \ref{fig5} we investigate the effect of 
anisotropy of the $\kappa$ indexes in the ion distribution, considering 
isotropic electron and ion temperatures.
The column at the left show the values of $z_i$, and the right column show
the values of $z_r$. 

In the case of figure \ref{fig4}, we consider a fixed value
of the relative dust density, $\epsilon= 2.5\times 10^{-6}$, and 
other parameters as in figure \ref{fig1}. For this figure, we assume PBK 
ion distributions, with isotropic temperatures,
$\kappa_{i\parallel}= 30$, and four values of $\kappa_{i\perp}$, namely
$\kappa_{i\perp}= 20$, 10, 5, and 2.93.  
These values of the ion kappa parameters correspond to effective ion 
temperature ratio given by 1.06, 1.19, 1.58, and 2.99, respectively.
Panels (a) and (b) depict results obtained considering isotropic Maxwellian
distribution for electrons, and panels (c) and (d)) show results obtained
considering PBK distribution for electrons, with $\kappa_{e\perp}
=\kappa_{e\parallel}=5.0$, and isotropic temperatures.
We can compare the results appearing in figure \ref{fig4}(a) with those
in figure \ref{fig1}(d), which was obtained considering the same parameters
and similar combination of velocity distribution functions, except that in
figure \ref{fig1}(d) the ion kappa indexes were isotropic and the ion 
temperatures were anisotropic.
For instance, it is seen in figure \ref{fig1}(d) that 
for $T_{i\perp}/T_{i\parallel}=3.0$
there is instability in the range $0.2\le q\le 1.4$, with maximum value 
$z_i\simeq 0.25$. In
the case of figure \ref{fig4}(a) the instability which appears for
$\kappa_{i\perp}= 2.93$ ($\theta_{i\perp}/\theta_{i\parallel}= 2.99$) 
occurs for $0.3\le q\le 0.9$, with
smaller value of the maximum growth rate. 
Another comparison which can be made is between the growth rates
appearing in figures \ref{fig4}(c) and \ref{fig1}(h). Both figures
show results obtained with PBK distributions for 
ions and electrons, with anisotropic temperatures in figure \ref{fig1} and
anisotropic ion kappa indexes in figure \ref{fig4}.
In figure \ref{fig1}(h) the instability for 
$T_{i\perp}/T_{i\parallel}=3.0$ occurs in the region 
$0.22\le q\le 1.15$, with maximum growth rate which approaches
$z_i=0.2$. In figure \ref{fig4}(c), the case with $\kappa_{i\perp}=2.93$,
for which the ratio of effective temperature is
$\theta_{i\perp}/\theta_{i\parallel}= 2.99$, does not show positive values
of $z_i$.
The conclusion which can be drawn is that the anisotropy in the effective
temperature which is due to the anisotropy of the kappa parameters of the
ion distribution is not so effective as the anisotropy of the ion
temperature parameters, on the
effectiveness of the EMIC instability.

In figure \ref{fig5} we investigate with more detail the role of the dust
density in the case of ions with a PBK distribution, considering 
several values of the dust relative density, and other parameters 
which are the same as in the case of figure \ref{fig1}.
Figures \ref{fig5}(a) and (b) show results obtained assuming a PBK ion 
distribution with isotropic temperatures and anisotropic kappa parameters, 
with an isotropic Maxwellian to describe the electron population. 
The ion kappa parameters are 
$\kappa_{i\parallel}= 30$ and $\kappa_{i\perp}= 2.93$, 
so that the effective temperature anisotropy of the ion distribution is 
$\theta_{i\perp}/\theta_{i\parallel}= 2.99$. 
For this case, the values of $z_i$ seen in figure \ref{fig5}(a) show that
the EMIC instability disappears for dust population between the values
$\epsilon= 2.5\times 10^{-6}$ and $\epsilon= 5.0\times 10^{-6}$. 
Figures \ref{fig5}(c) and \ref{fig5}(d) show results obtained assuming
that the electron distribution function is a PBK distribution, with isotropic
temperatures and $\kappa_{e\parallel}=\kappa_{e\perp}= 5.0$, instead of an 
isotropic Maxwellian, with the ion distribution and other parameters equal
to those used in the case shown
in figures \ref{fig5}(a) and \ref{fig5}(b).
This example shows that, also in the case of ion distribution with anisotropy
due to the kappa indexes, the increase of the non thermal character of the
electron distribution, with the presence of a power-law electron tail, 
contributes to decrease the EMIC instability due to the anisotropic ion
distribution.

For another comparison with the results shown in figure
\ref{fig5}(a), figure \ref{fig5}(e) shows the values of $z_i$ obtained 
with isotropic Maxwellian distribution for electrons and PBK distribution
for ions, with isotropic ion kappa indexes 
$\kappa_{i\perp}=\kappa_{i\parallel}= 5.0$ and anisotropy due to the ion 
temperatures. We choose the ion temperature ratio as 
$T_{i\perp}/T_{i\parallel}= 2.565$, which leads to the
same anisotropy of the effective temperatures as in figure \ref{fig5}(a).
It is seen that in the case of figure \ref{fig5}(e) the instability is 
slightly stronger, with larger range of unstable wave numbers, than in the
case of figure \ref{fig5}(a). As already noticed in the commentaries about
figure \ref{fig4}, it is seen that the anisotropy which is due to anisotropic
ion kappa indexes is not so effective in producing the EMIC instability as the
anisotropy due to the ion temperatures.

\section{Conclusions}
\label{sec:conclusions}

We have presented results obtained from a numerical analysis of the 
dispersion relation for low-frequency
ion-cyclotron waves propagating along the ambient
magnetic field in a dusty plasma,
considering situations where the velocity
dispersion along perpendicular direction is greater than along the parallel
direction, i.e., considering situations which can lead to the ion-cyclotron
instability. It has been considered that
either ions or electrons, or both, can have product-bi-kappa velocity 
distributions.
Regarding the influence of the shape of the velocity
distribution, results obtained considering absence of dust have shown that, 
if the electrons are described by an isotropic Maxwellian distribution
and the ion distribution is changed from the bi-Maxwellian limit to a PBK
distribution with relatively small kappa index, the range in wavenumber
where instability occurs is increased, and the magnitude of the growth rates
is increased as well. Conversely, if the ion distribution is a bi-Maxwellian
distribution and the electron distribution is changed from
a Maxwellian distribution into a PBK distribution
with small kappa index, with isotropic electron temperature, the magnitude of
the growth rates and range of the EMIC instability is somewhat reduced, but
the magnitude of the reduction is less significant than the increase due to
ion PBK distributions. 
Regarding the influence of dust, the results obtained
have shown that, for all shapes of ion and
electron distributions which have been utilized, 
the presence of a small population
of dust leads to some decrease in the magnitude of the growth rates and in the
range of the 
EMIC instability, in comparison with the case without dust. We have also seen
that in presence of dust a region of wave damping appears 
for small values of $q$, which is not
present in the dustless case. The appearing of this
damping for large wavelengths, and the overall reduction in the growth rates
of the instability in the presence of dust, are consequence of the collisional
charging of the dust particles. 
All these features, related to different forms
of ion and electron distributions and to the presence of a small amount of
dust, have been obtained considering the plasma parameter $\beta_i=2.0$, 
and have also been verified considering one example with smaller value of this
parameter, $\beta_i=0.2$.

Still regarding the effect of the presence of dust, the results obtained have 
shown that a value of the relative number density of dust
which is enough to make disappear
completely the EMIC instability in bi-Maxwellian plasma, may be not enough to
overcome the instability in plasmas with ions described by PBK distributions
with small kappa index. It has also been seen that the increase in the 
dust population leads to decrease of the phase velocity of the IC waves, for
all types of combinations of Maxwellian and PBK distributions which have been
investigated.

For fixed temperature ratio $T_{i\perp}/T_{i\parallel}$ and fixed dust number 
density, it has been seen that
the decrease of the kappa index in a PBK ion distribution leads to increase of
the instability in range and in magnitude of the growth rate, and that 
the increase becomes more and more pronounced for progressively smaller ion
kappa indexes. On the other hand, for fixed value of 
$T_{i\perp}/T_{i\parallel}$ and dust density ratio, with bi-Maxwellian
distribution for ions, the decrease of the kappa index in a PBK electron
distribution leads to decrease of the EMIC instability, in magnitude of the
growth rate and in range,
but the decrease is not very pronounced. 
These features can be qualitatively
explained as follows. For low frequency waves, the real parts of the 
arguments of the modified $Z$ function, which appear in the dispersion 
relation, are such that $|\zeta_i^s/\zeta_e^s|= \sqrt{(m_i/me)
(T_{i\parallel}/T_{e\parallel})}$. In the case of bi-Maxwellian ion 
distribution and Maxwellian electron distribution, with $T_{i\parallel}
=T_{e\parallel}$, it is known that the ion contribution is much more 
significant than the electron contribution, for the EMIC instability. 
For PBK ion distribution, the population of
high energy ions is increased in comparison with the Maxwellian case, and can
be compared to the high energy population of a higher temperature
thermal plasma. Therefore, the instability growth rates of the EMIC
are increased with the increase of the non thermal character of the PBK ion
distribution. On the other hand, if the electrons are described by a PBK
distribution, the population of resonant electrons may be much increased,
in comparison with the thermal case. It is as if in the resonant region the
ratio $|\zeta_i^s/\zeta_e^s|$ is decreased, by increase of an effective
$T_{e\parallel}$. Therefore, electron cyclotron damping sets in, and the 
EMIC growth rates are reduced in comparison with the thermal case.

We have also investigated the effect of anisotropy in the kappa indexes
of the ion PBK distribution, for given dust number density and for isotropic
ion temperatures. It has been seen that the anisotropy due to the anisotropic
kappa indexes is not so effective in producing the EMIC instability as the
anisotropy due to the ratio between the ion perpendicular and parallel 
temperatures.

%% Acknowledgements
%
% \acknowledgments
\acknowledgments
MSdS acknowledges support from Brazilian agency CAPES.
LFZ acknowledges support from CNPq (Brazil), grant No. 304363/2014-6.
RG acknowledges support from CNPq (Brazil), grants No. 304461/2012-1 and
478728/2012-3. 
% <Acnowledgments text>

%% References
%% Please cite all reference entries in the article text using \cite or
%% equivalent command. 

%%%  Using BibTeX  (Name-Year style)
%
%\bibliographystyle{spr-mp-nameyear-cnd}  %% BibTeX style
% \bibliographystyle{/home/ziebell/zb_2/tex_other/spr-mp-nameyear-cnd}  %% BibTeX style
% \bibliography{<bib data>}                %% BibTeX data
%\bibliography{/home/ziebell/zb/bib/refs,/home/ziebell/zb/bib/livros,%
%/home/ziebell/zb/bib/plart,/home/ziebell/zb/bib/plconf,%
%/home/ziebell/zb/bib/plphd,/home/ziebell/zb/bib/plmsc,%
%/home/ziebell/zb/bib/plsub}

%% Non-BibTeX  (Name-Year style)
%
% \begin{thebibliography}{}
% \bibitem[\protect\citeauthoryear{<author>}{<year>]{ref:?}
%    <ref. entry>
% \bibitem[\protect\citeauthoryear{<author>}{<year>]{ref:?}
%    <ref. entry>
% \end{thebibliography}

\end{document}